# Regularization strategy for the layered inversion of airborne TEM data: application to VTEM data acquired over the basin of Franceville (Gabon).

Julien Guillemoteau[1], Pascal Sailhac[1] and Mickaël Béhaegel[2]

[1] *Institut de Physique du Globe de Strasbourg & EOST, CNRS-UDS UMR 75-16, Strasbourg, France.*

[2]*Areva NC, Geoscience Direction, Mining Business Group, La Défense, Paris, France.*

Corresponding author*: j.guillemoteau@unistra.fr*

**Abstract**

Airborne transient electromagnetic (TEM) is a cost-effective method to image the distribution of electrical conductivity in the ground. We consider layered earth inversion to interpret large data sets of hundreds of kilometre. Different strategies can be used to solve this inverse problem. This consists in managing the *a priori* information to avoid the mathematical instability and provide the most plausible model of conductivity in depth.

In order to obtain fast and realistic inversion program, we tested three kinds of regularization: two are based on standard Tikhonov procedure which consist in minimizing not only the data misfit function but a balanced optimization function with additional terms constraining the lateral and the vertical smoothness of the conductivity; another kind of regularization is based on reducing the condition number of the kernel by changing the layout of layers before minimizing the data misfit function. Finally, in order to get a more realistic distribution of conductivity, notably by removing negative conductivity values, we suggest an additional recursive filter based upon the inversion of the logarithm of the conductivity.

All these methods are tested on synthetic and real data sets. Synthetic data have been calculated by 2.5D modelling; they are used to demonstrate that these methods provide equivalent quality in terms of data misfit and accuracy of the resulting image; the limit essentially comes on special targets with sharp 2D geometries. The real data case is from Helicopter-borne TEM data acquired in the basin of Franceville (Gabon) where borehole conductivity loggings are used to show the good accuracy of the inverted models in most areas, and some biased depths in areas where strong lateral changes may occur.

## Introduction

Airborne transient electromagnetic (TEM) surveying was introduced about fifty years ago in the mining industry to detect shallow conductive targets like graphitic or sulphide formations. Nowadays, this method is also useful for groundwater exploration (Auken *et al*., 2009) or on-shore hydrocarbon exploration (Huang and Rudd, 2008).

Thanks to recent improvement in acquisition systems, it is now possible to image continuously, quickly and accurately the electrical conductivity distribution in the ground with the development of new modelling and inversion strategies. 2D inversion (Wolfgram *et al*, 2002), 2.5D inversion (Wilson and Raiche, 2006) or 3D inversion (Cox *et al*, 2010) starts to become practical when applied on airborne electromagnetic (AEM) data. However, less accurate interpretation as layered earth inversion remains the most useful method to interpret fastly large amount of data or to provide prior model for fast 3D inversion.

The first step of this method is to carefully define the 1D kernel relating the model of conductivity in depth to the data of apparent conductivity in time or frequency, the second step is to carefully invert the data. Actually in most cases, the layered inversion is an ill-posed problem which needs regularization. Zhdanov (2009) provided a detailed description of the recent improvement notably the minimum support method (Portniaguine and Zhdanov, 1999) concerning regularization problem in EM geophysics.

For 1D AEM inverse problem, one can use standard Tikhonov strategy. Christensen (2002) developed a fast method called “One Pass Imaging” which uses a regularization of the *z*-variability (in the vertical direction). Siemon *et al.* (2009) and Vallée and Smith (2009) recently published other results obtained by regularization with horizontal constraints. In complement to all these approaches, in this paper we expose and compare three others methods to solve the 1D inverse problem of AEM data.

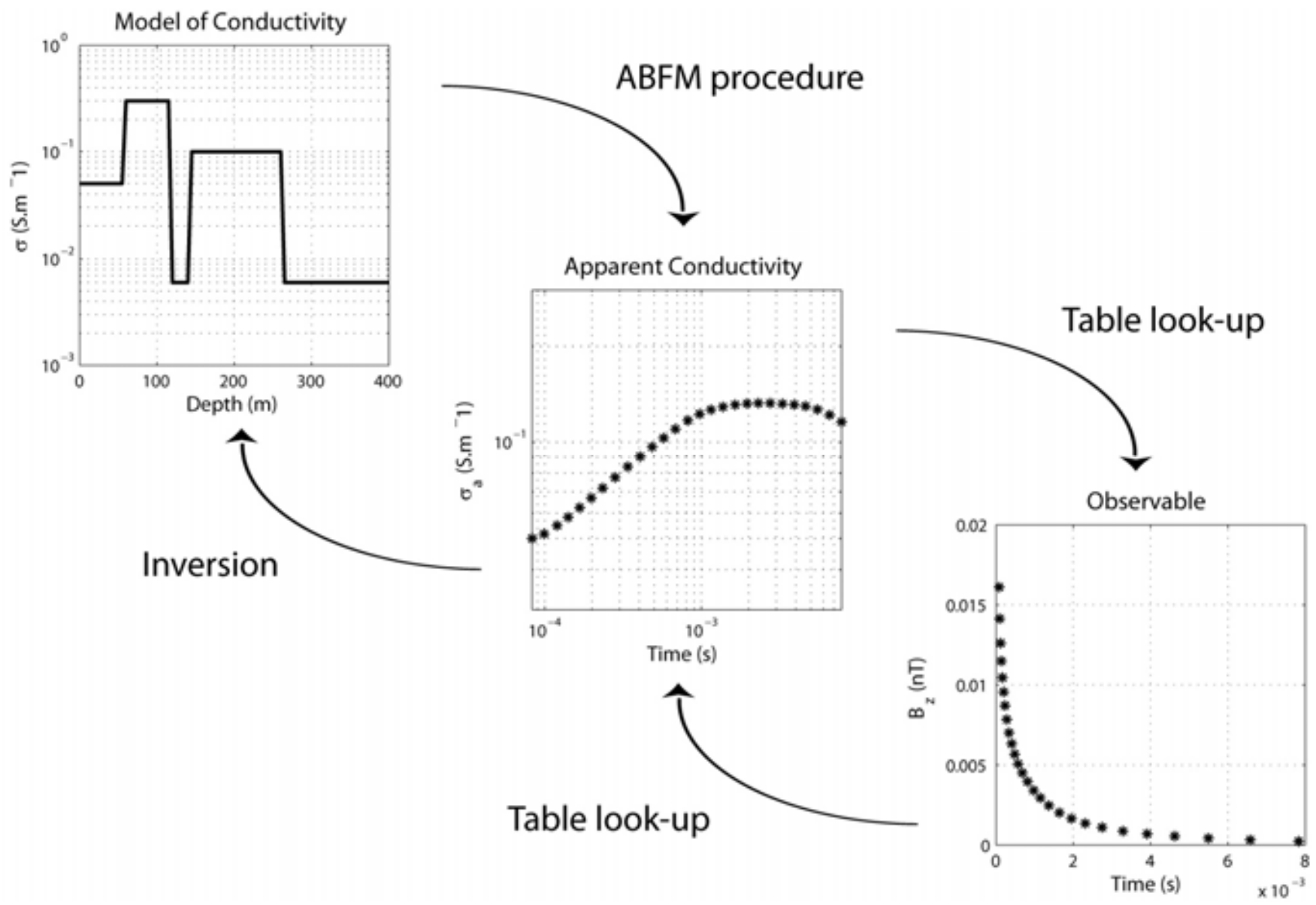

**Figure 1:** *Description of the procedure for the interpretation of TEM data. The definition of the apparent conductivity aims to avoid the configuration dependant part of the problem. The forward modelling consists in computing first the apparent conductivity by using the ABFM procedure. Secondly, the TEM response is found out by table look up of the homogeneous response which has been previously computed using relation (1) for large number of conductivities. The inversion of TEM data is the reciprocal process.*

The first one uses a constraint over the vertical derivatives and is similar to the One Pass Imaging developed by Christensen (2002). As AEM data are over-sampled along the flight line direction, we developed a second approach which uses this information to apply lateral constraint. This method is based upon a minimum length criterion over the difference with the results from the previous sounding. We suggest a third new approach which allows getting a natural generalized inverse (which mean no regularization) and which is based upon local analysis of the condition number to determine the layer layout prior to the inversion. All our programs use the linear modelling called Adaptative Born Forward Mapping (ABFM) to predict the data (Christensen, 2002). The ABFM procedure consists in solving a linear relationship between the apparent conductivity and the real conductivity. It is based on the hypothesis of normal distribution of conductivities, which is able to result in a model with some negative values; the better physical hypothesis which constrains any conductivity to be positive is that they obey lognormal statistics. Thus, we suggest an additional step using the kernel for the logarithm of the conductivities; this avoids negative conductivity values in the resulting model and allows sharper boundaries within the resulting model.

First we compare these methods when applied on a simple synthetic case. The artificial measurements have been generated by 2.5D modelling using the program “ArjunAir_705” developed by the P223 EM modelling project (Raich, 2008, Wilson *et al.*, 2006). The quality of each method is considered in terms of error (data misfit) and comparison to the true model. Then, we apply these methods on real data set acquired over the basin of Franceville (Gabon) in order to detect the bottom of an ampelite layer characterized by relatively high electrical conductivities. In that real data case, the quality of each method is considered in terms of data misfit and comparison to borehole measurements.

## Imaging the electrical conductivity

*Description of the problem*

Airborne TEM data are provided in terms of magnetic field $h(t)$ or its time derivative $dh/dt$ recorded in the receiver loop, where $t$ is the time delay after turn off of the transmitter loop. In this paper, we consider the vertical magnetic field located at the centre of a horizontal circular loop transmitter; in the quasi static domain it is given in the Fourier domain by (Ward and Hohmann, 1987):

$$H_z(\omega) = \frac{Ia}{2}\int_0^{\infty}\left[e^{-\lambda z} + r_{TE}e^{\lambda z}\right]\lambda J_1(\lambda a)d\lambda, \qquad (1)$$

where $a$ is the radius of the transmitter loop, $z$ is its altitude and $I$ is the amplitude of the electrical current injected, $\lambda$ is the horizontal component of the wave number in the air and $r_{TE}$ is the reflection coefficient which depends on the conductivity of the underground medium. Because the typical transmitter current is a step current with turn off at time 0, the transient response in terms of $dh_z/dt$ is computed by performing the inverse Laplace transform of the expression (1). Then, $h_z(t)$ can be obtained by integrating $dh_z/dt$. Finally the response of the system is given by convolving the step response to the time derivative of the actual current injected in the transmitter loop.

One way to simplify the problem is first to convert $h_z(t)$ data into apparent conductivity and then to compute the layer conductivities by inversion of the apparent conductivities. The apparent conductivity $\sigma_a$ is defined as the conductivity of the equivalent homogeneous half space which provides the same response. Therefore, $\sigma_a$ is the solution of the following equality:

$$h_z^{half\ space}(t,\sigma_a) = h_z^{tabular}(t,\{\sigma_i,h_i\}). \qquad (2)$$

By introducing the apparent conductivity, one can separate the problem into two subsequent parts: 1- one configuration dependent part which is the relation between the TEM response and the apparent conductivity characterised by Relation (1) for a homogeneous half space, 2- a configuration independent part which relies the apparent conductivity to the layer conductivities. To interpret our data, we follow the reciprocal scheme (see Figure 1). For each time window, the apparent conductivity is inverted by table look-up within abacus containing the current system response $h_z(t)$ or $dh_z/dt$ for a large amount of homogeneous half space. To compute the real conductivities, we use the ABFM method (Christensen, 2002) which is the linearized version of Equation (2) in time-domain. For a layered medium with $N_m$ layers, the apparent conductivity versus the $N_d$ time windows is written as a linear combination of the conductivity versus depth:

$$\sigma_{a,i} = \sum_j F_{ij}.\sigma_j \qquad i = 1, N_d \quad j = 1, N_m, \qquad (3)$$

where $\boldsymbol{\sigma_a}$ is the vector of apparent conductivity, $\boldsymbol{\sigma}$ is the vector of the layer conductivities and $\mathbf{F}$ is the kernel depending on the apparent conductivity and time. The latter relation constitutes the forward formulation of our problem. Assuming that the measured response is always above noise level, the maximum depth of the layered media is given equal to penetration depth of the primary field at the largest time window. This depth can be approximated by the following relationship introduced by Christensen (2002):

$$z_{max} = \sqrt{\frac{ct_{N_d}}{\sigma_{a,N_d}\,\mu_0}}, \qquad (4)$$

where *c=2.8* is the ad hoc scaling factor that he obtained by minimizing the squared difference between the exact and the approximate apparent conductivities (Equation 3) summed over six layered models with different parameters.

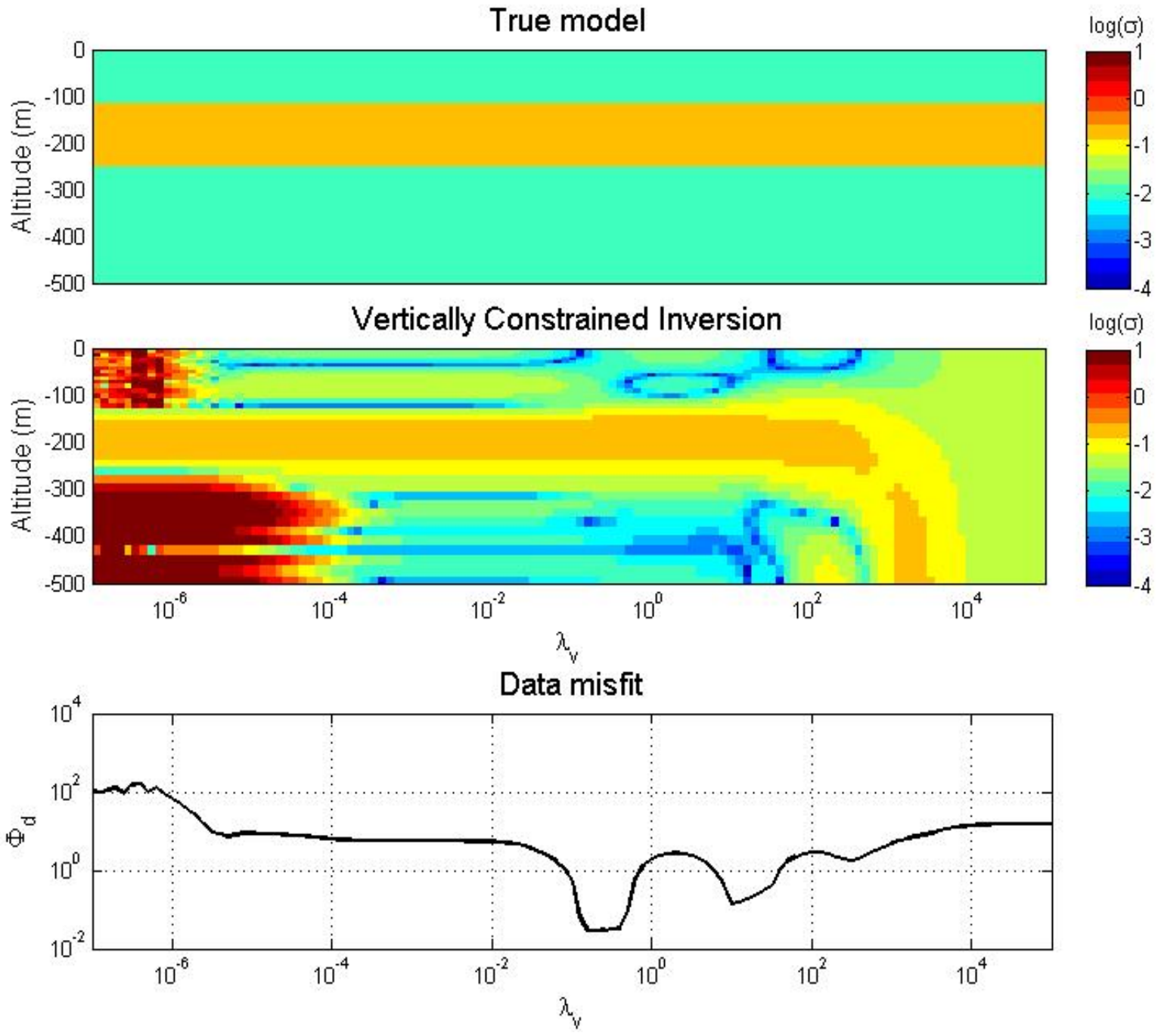

**Figure 2:** *Vertically constrained inversion of synthetic noisy data (0.1% of random noise added to $h(t)$) for a large number of $\lambda_V$. The data misfit versus regularization describes the balance between data information and constraints based on a priori information.*

In order to follow the decrease of resolution with depth, we set the grid interfaces in such a way that the $n^{th}$ layer is *n* times thicker than the first one. Depending on the magnitude of the discretization of the media, the problem is over determined, mixed determined or under determined. In most cases, we consider mixed-determined problems because we need a compromise when choosing the number of layers: enough layers are necessary to get a good vertical resolution, however too much layers would increase the computation time above acceptable values for real time applications. In the following, we discuss the different strategies to solve such a 1D inverse problem.

*Inversion with vertical constraints*

The vertical constraint has been suggested to regularize 1D TEM problem by Christensen (2002) in counterpart to the measure of the length of the solution. In this study, we consider only the vertical constraint in order to identify its own effect. The objective function which has to be minimized is given as follows:

$$\Phi = \sum_{i=1}^{N_d} e_i^2\left(\sigma_{a,i} - F_i(\boldsymbol{\sigma})\right)^2 + \lambda_V{}^2\left[\left(\frac{d\sigma_1}{dz}\right)^2 + \sum_{j=2}^{N_m-1}\left(\frac{d^2\sigma_j}{dz^2}\right)^2 + \left(\frac{d\sigma_{N_m}}{dz}\right)^2\right] \quad (5)$$

The first sum of this expression constitutes the data misfit (*e.g.* a weighted least square) and the second constitute the regularizing part. $e$ is a weighting vector characterizing the importance of each measurement, it could be related to the inverse of variances if one also considers independent normal distributions for the apparent conductivity at each time delay $t_i$; $\lambda_V$ is a weighting factor characterising the vertical variability of the model. The solution which minimizes the function $\Phi$ is written as follows (Menke, 1989):

$$\boldsymbol{\sigma} = [\mathbf{F}^T\mathbf{W_d}\mathbf{F} + \mathbf{S}]^{-1}\mathbf{F}^T\mathbf{W_d}\boldsymbol{\sigma}_a \ , \quad (6)$$

where $\mathbf{W_d}$ is a diagonal matrix containing the weighting factors $e_i^2$. We took $\mathbf{W_d}$ equal to an identity matrix for all the inversions discussed in this paper since we do not have any *a priori* information on the measurements. $\mathbf{S} = \lambda_V{}^2\mathbf{D}^T\mathbf{D}$ is the vertical smoothness matrix which depends on the first and second order derivative of the model, $\mathbf{D}$ is written as:

$$\mathbf{D} = \begin{bmatrix} 1/\delta z_1 & -1/\delta z_1 & & & \\ 1/\delta z_2^2 & -2/\delta z_2^2 & 1/\delta z_2^2 & & \\ & & \ddots & & \\ & & 1/\delta z_{N_m-1}^2 & -2/\delta z_{N_m-1}^2 & 1/\delta z_{N_m-1}^2 \\ & & & -1/\delta z_{N_m} & 1/\delta z_{N_m} \end{bmatrix}, (7)$$

where $\delta z_i$ is equal to half of the current layer thickness in order to compensate the increasing thickness of layers with depth. Figure 2 shows the inversion and the relative apparent conductivity misfit of 1D synthetic data for an increasing smoothness. We tested values of regularization control parameter $\lambda_V$ in the range of $[10^{-7}, 10^5]$. In the bottom of Figure 2, the apparent conductivity misfit is displayed versus the regularization weight. If $\lambda_V$ is too small, the regularization is two weak and the matrix to be inverted is singular. If $\lambda_V$ increases, the smoothness becomes more important: this avoids artefact due to the noise of data until an optimal value $\lambda_V \sim 0.5$ which well reproduces the initial model. For larger $\lambda_V$, the regularization becomes preponderant and covers the information contained in the data: the smoothness is too large to reproduce real conductivity changes and data errors increase as well.

Different approaches exist to find the optimal value for lambda (Hansen, 1992, 2010) or (Oldenburg and Li, 1994; Constable and Parker, 1987) for iterative methods. We used the simplest criterion similar to the discrepancy principle (Aster *et al.*, 2005): we set the optimal parameter at the highest value of $\lambda_V$ able to produce a reasonable data misfit. Consequently, one has to define the threshold of the data misfit regarding the level of noise before the inversion.

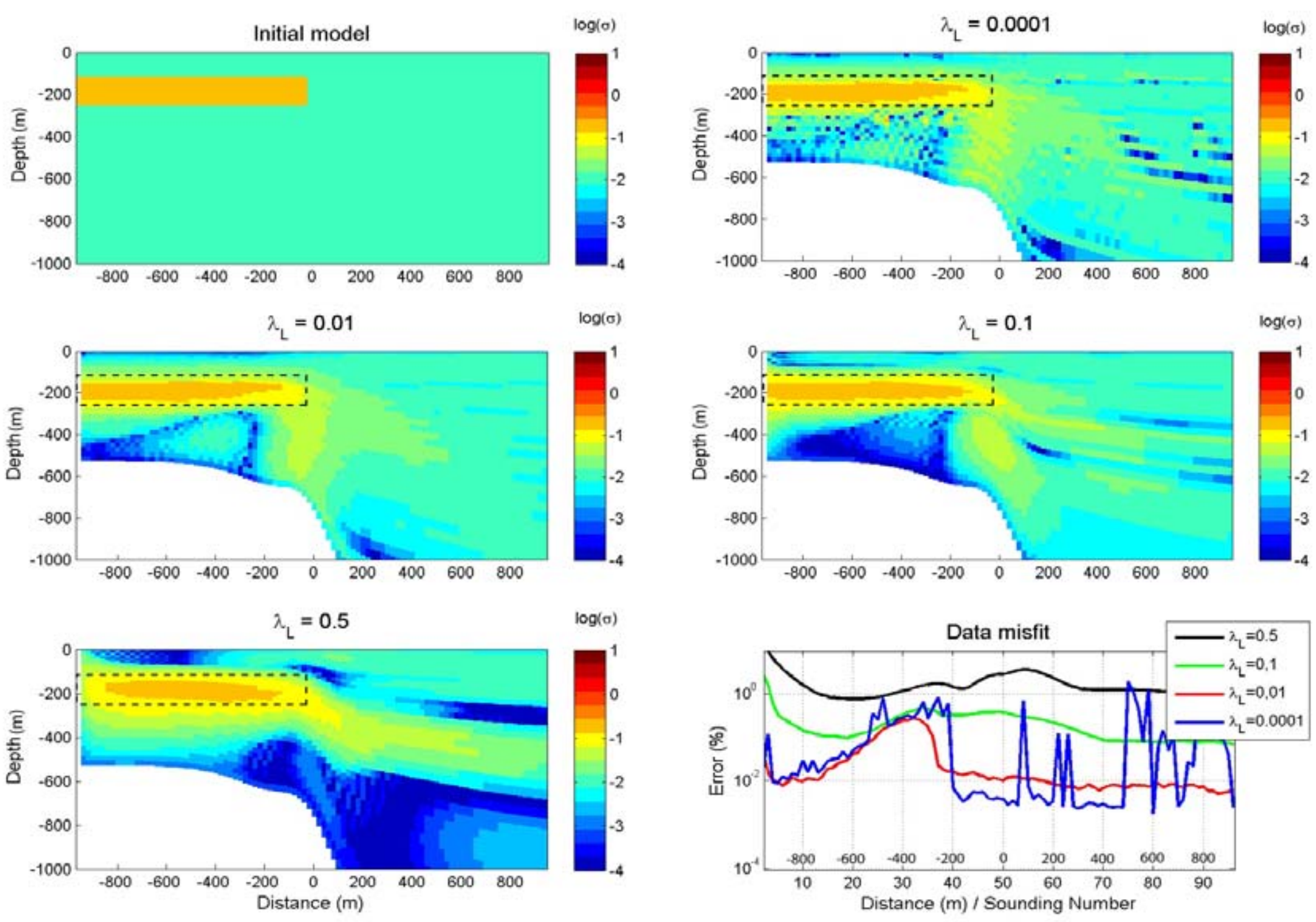

**Figure 3:** *Laterally constrained inversion of noisy (0.1%) 2.5D synthetic data for different values of* $\lambda_L$*. The optimal choice of regularization is around* $\lambda_L = 0.01$*.*

*Inversion with lateral constraints*

Airborne TEM are usually over sampled along the flight line. Indeed, since the footprint of the method is larger than the interval between two soundings, the measurements cannot vary sharply. It is possible to use this information as a lateral constraint to regularize this 1D inverse problem. Monteiro Santos *et al.* (2004) for ground measurements, Auken *et al.* (2005) and Viezzolli *et al.* (2008) for airborne data set, developed 1D Laterally Constrained Inversion (LCI) based on a smoothing term which constraints lateral derivatives of the model. In these methods, one needs to consider several soundings simultaneously. Christiansen *et al*. (2007) applied this method on small data set from ground-based measurements. Siemon *et al*. (2009) adapted this method for large airborne continuous measurements. In order to reduce the computational cost, we propose a method which allows inverting all the soundings independently with a simple lateral constraint. This method consists in using the result provided by the previous sounding as a reference model. In that case, the objective function that we have to minimize is composed of the data misfit plus a second term which minimizes the difference between the solution and the result provided by the previous sounding $\boldsymbol{\sigma_{s-1}}$:

$$\Phi = \sum_{i=1}^{N_d} e_i^2\left(\sigma_{a,i} - F_i(\boldsymbol{\sigma_s})\right)^2 + {\lambda_L}^2 \sum_{j=1}^{N_m}\left(\sigma_{s,j} - \sigma_{s-1,j}\right)^2. \quad (8)$$

The solution $\boldsymbol{\sigma_s}$ of the current sounding is given by:

$$\boldsymbol{\sigma_s} = \boldsymbol{\sigma_{s-1}} + [\mathbf{F^T W_d F} + {\lambda_L}^2\mathbf{I}]^{-1}\mathbf{F^T W_d}[\boldsymbol{\sigma_a} - \mathbf{F}\boldsymbol{\sigma_{s-1}}]. \quad (9)$$

The parameter $\lambda_L$ controls the magnitude of the lateral constraint. Figure 3 shows the laterally constrained inversions of 96 synthetic *Bz* data with 0.1% of noise performed every 20 meters over a 2.5D conductivity model. The synthetic data was generated by using the program ArjunAir with which we simulate pure step response in coincident loop geometry (with a loop transmitter of 26m in diameter and a 1.1m diameter receiver and a nominal clearance of 45m). They are composed of 27 channels starting from $t_1 = 83\,\mu s$ to $t_2 = 7.8\,ms$ . The aim is to reproduce VTEM (Witherly *et al.*, 2004) configuration. The model is composed of a conductive layer of conductivity σ=200mS/m embedded within a host medium of 10 mS/m. The results of the inversion and their relative misfit are displayed in Figure

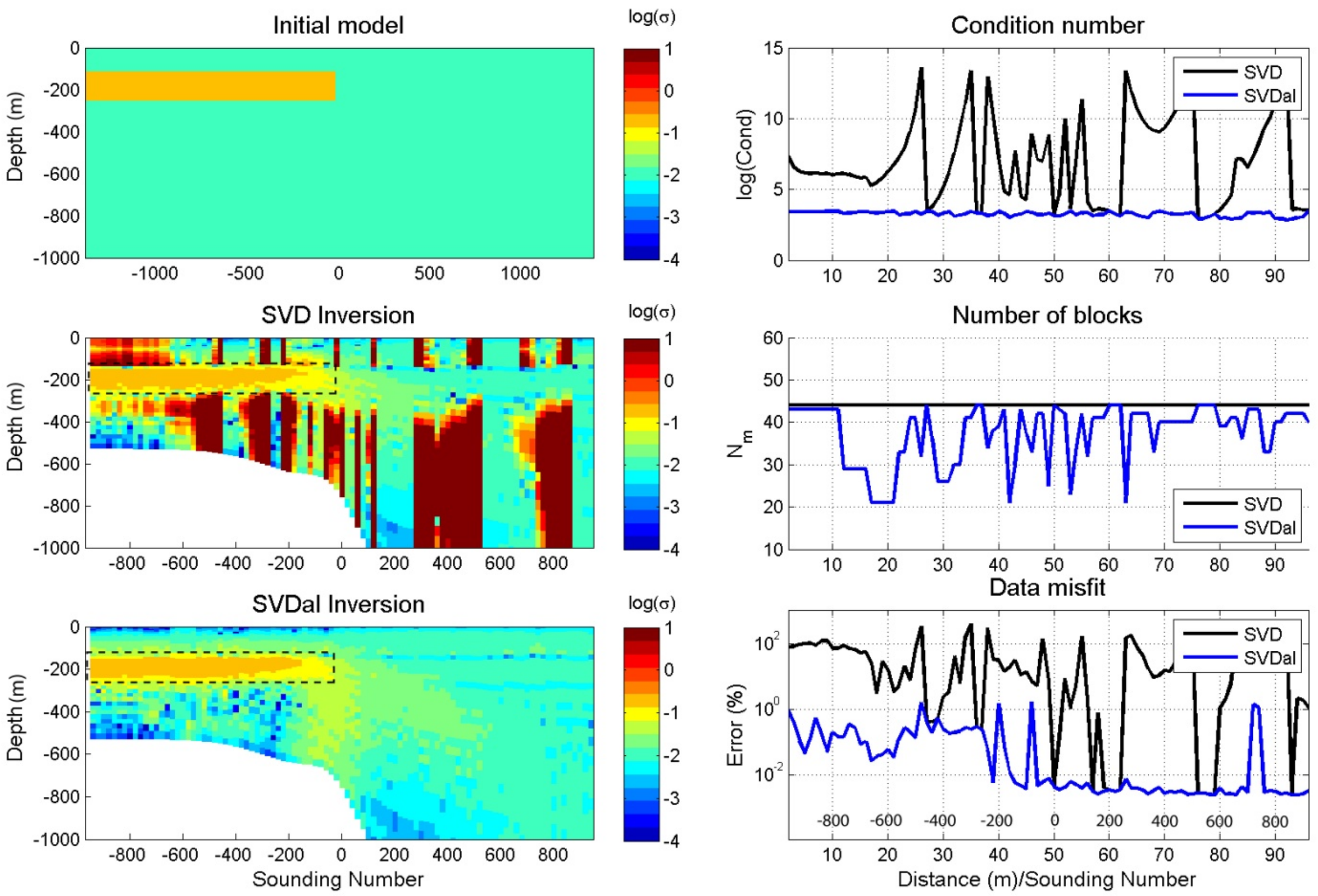

**Figure 4:** *Comparison between simple SVD inversion and SVDal inversion of a noisy (0.1%) 2.5D synthetic dataset. The algorithm SVDal changes the number of layers in order to reduce the condition number and avoid singular values.*

3 for four increasing degrees of smoothness ($\lambda_L = [0.001, 0.01, 0.1, 0.5]$). Like in the previous section, setting $\lambda_L$ too small allows non realistic values of conductivities generating unstable data misfit. By comparing to the true conductivity, we can conclude that the optimal choice of the regularization factor is around $\lambda_L = 0.01$. For larger smoothness, the inversions need more soundings to converge toward the right model. If a strong lateral variation occurs, the data misfit increases first and then decreases as slow as $\lambda_L$ is large. We conclude that $\lambda_L$ has to be chosen reasonably after the consideration of the lateral data sampling. Indeed, if the sampling rate increases, larger value of $\lambda_L$ can be efficient since the lateral influence will be more important.

*SVD inversion with adaptative layout 'SVDal'*

Another way to solve an inverse problem is to use the natural generalised inverse provided by singular value decomposition (Lanczos, 1961):

$$\boldsymbol{\sigma} = \mathbf{V_p}\boldsymbol{\Lambda}_\mathbf{p}^{-1}\mathbf{U}_\mathbf{p}^\mathbf{T}\boldsymbol{\sigma}_\mathbf{a}\,, \tag{10}$$

where $\mathbf{V_p}$ and $\mathbf{U_p}$ are the matrices of the $p$ eigenvectors related to non-null eigenvalues and spanning the model space and the data space respectively. The advantage of this method is that we do not need any *a priori* information. Thus, the natural generalized inverse leads to minimize only the term of data misfit without any weighting factors:

$$\Phi = \sum_{i=1}^{N_d}\left(\sigma_{a,i} - F_i(\boldsymbol{\sigma})\right)^2 . \tag{11}$$

However, for realistic cases which are ill-conditioned problems, the eigenvalues smoothly decrease toward zero so that it is difficult to identify non-null ones. In practice, the solution is to cut off the small eigenvalues or to damp them like Huang and Palacky (1991) or Chen and Raiche, (1998). Actually, this method is equivalent to a least square inversion with a weighted constraint on the length of the solution. In order to keep a natural solution (which minimizes only the term of data misfit), we propose a method which consists in designing the grid of the model before the inversion in such a way that the problem is well conditioned. The algorithm of

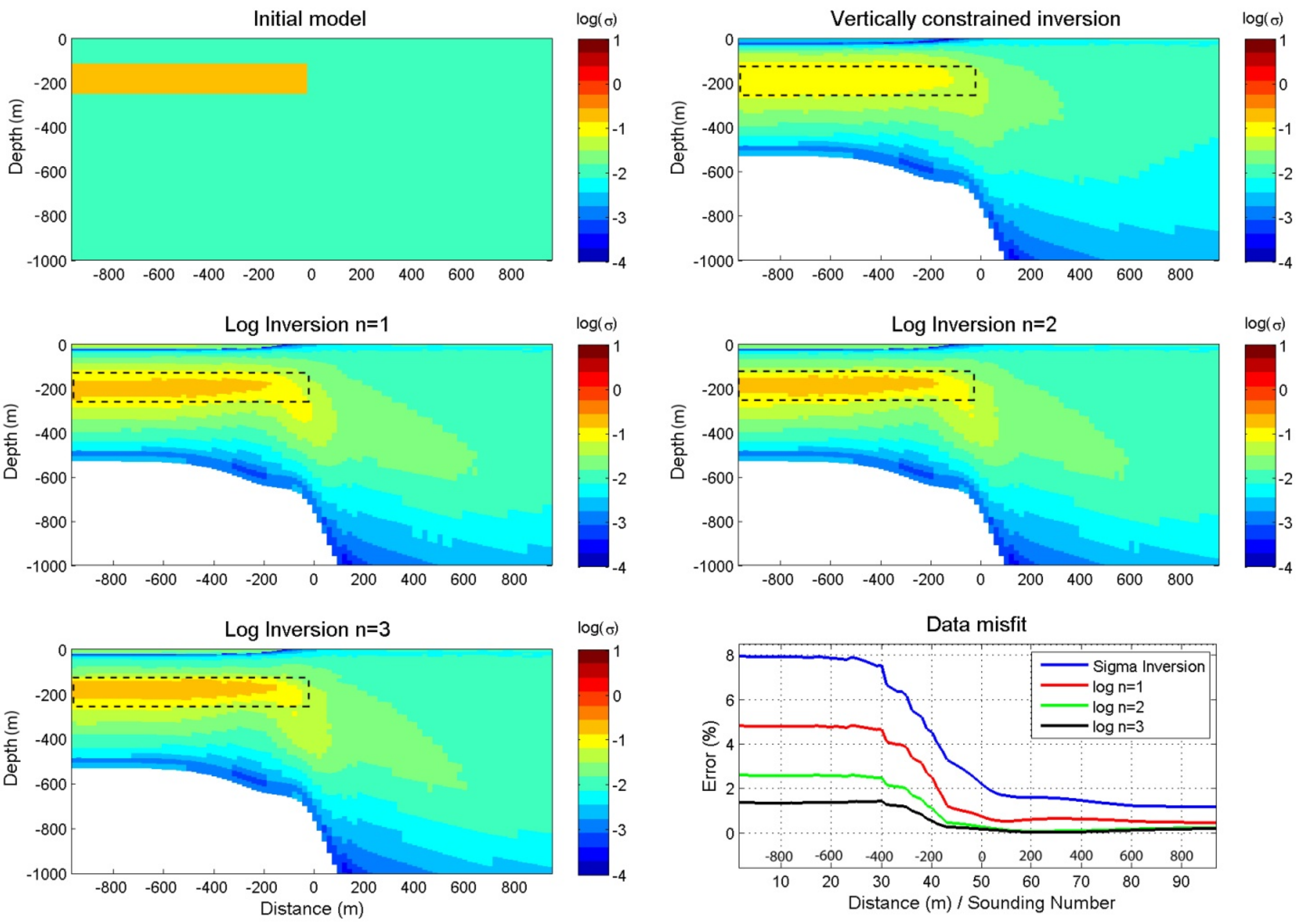

**Figure 5:** *Application of the logarithmic inversion to a synthetic data set. The starting model which has been used is a smooth result provided by the vertically constrained inversion. The results show a good convergence in term of resulting image and data misfit.*

this method which we call SVDal can be described as follows:

1- Knowing the maximum depth investigation, we compute the grid layout for a initial number of layers.
2- We compute the condition number which is defined as the ratio between the highest and the lowest non-zero eigenvalues of the kernel.
3- If the condition number is too high, we change the number of layers and execute again the previous steps until the condition number is sufficiently low.
4- At last, the solution is computed using the relation (10).

Figure 4 shows the difference between a simple SVD inversion with 45 layers and the SVDal inversion. For simple SVD inversion, some soundings may be ill-conditioned; the singularities generate infinite values of conductivity. These strong artefacts are correlated with large condition numbers. By changing the number of layers in the SVDal algorithm, one reduces these large condition numbers to a more reasonable value which has been fixed to $Cond = 3000$ before the SVD inversion. The application on synthetic data shows a good misfit associated to the right convergence in the model space. As the condition number decreases naturally with the numbers of unknowns, it is important to understand that the SVDal algorithm do not find the lowest value of the condition number but the nearest reasonable one. By choosing a decreasing number of layers, one reduces obviously the computational cost and, in a way, raises the smoothness of the resulting model.

**Imaging the logarithm of the conductivity**

The principal disadvantage of the methods presented above is that they are fundamentally based on the assumption that conductivity is normally distributed: however it is well known that conductivity of rocks usually follows a log normal distribution (Palacky, 1987). In order to remove negative values and to get a more realistic distribution it is therefore more convenient to write the problem with the logarithm of the conductivity:

$$log\ (\sigma_{a,i}) = \sum_j F^*_{ij}.log\ (\sigma_j) \quad i = 1, N_d \quad j = 1, N_m, (12)$$

with

$$F^*_{ij} = \frac{\sigma_j}{\sigma_{a,i}} F_{ij} \quad . \qquad (13)$$

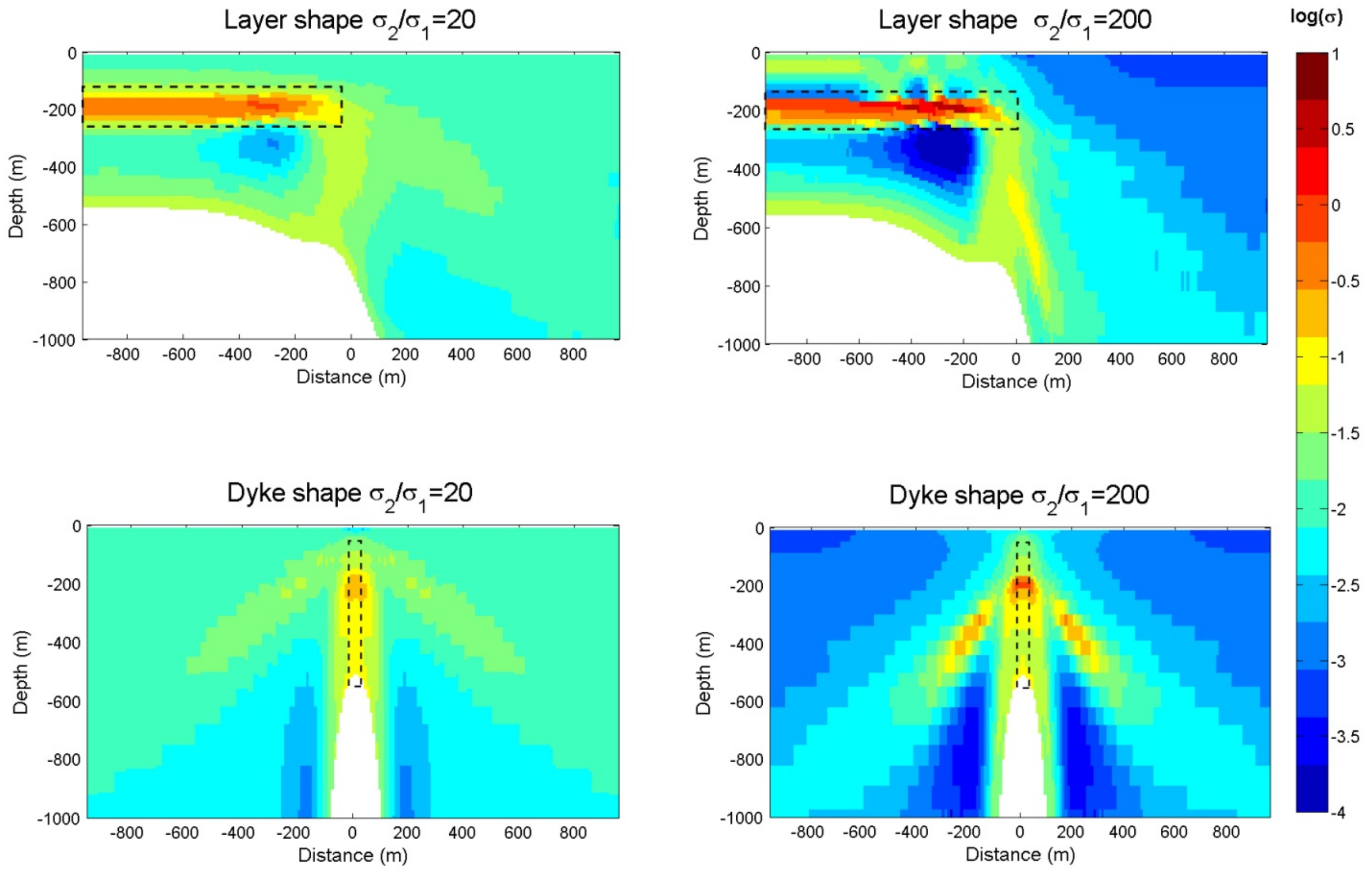

**Figure 6:** *Layered inversion of layer-shaped and dyke-shaped 2D conductive target within two different hosting media. One can see that 1D inversion fails if the conductivity contrast is higher or in case of vertically shaped formations.*

The new kernel $F^*$ is highly non linear because it depends on the model explicitly. Therefore, the problem has to be solved by using a non linear iterative method. If one knows a model being relatively close to the solution, one can use the perturbation theory. We can write the relation as follows:

$$log\ \sigma_{a,i} = log\ \sigma^0_{a,i} + \sum_j F^*_{ij}\left(log\ \sigma_j - log\ \sigma^0_j\right). \qquad (14)$$

By setting $y_{a,i} = log\ (\sigma_{a,i}/\sigma^0_{a,i})$ and $y_j = log(\sigma_j/\sigma^0_j)$, this relation leads back to a new formulation of the linear relation to be inversed:

$$y_{a,i} = \sum_j F^*_{ij} y_j \quad i = 1, N_d \quad j = 1, N_m \quad . \qquad (15)$$

This linear inverse problem is still partly undetermined and needs regularization. We can use the fact that the Taylor approximation allows only small perturbations. The level of perturbation can be regularized by minimising the length of the vector solution $Y$, so the objective function can be written as follows:

$$\Phi = \sum_{i=1}^{N_d} \left(y_{a,i} - \mathbf{F}^*_{\mathbf{i}}\mathbf{y}\right)^2 + {\lambda_P}^2 \sum_{j=1}^{N_m} {y_j}^2 \ . \qquad (16)$$

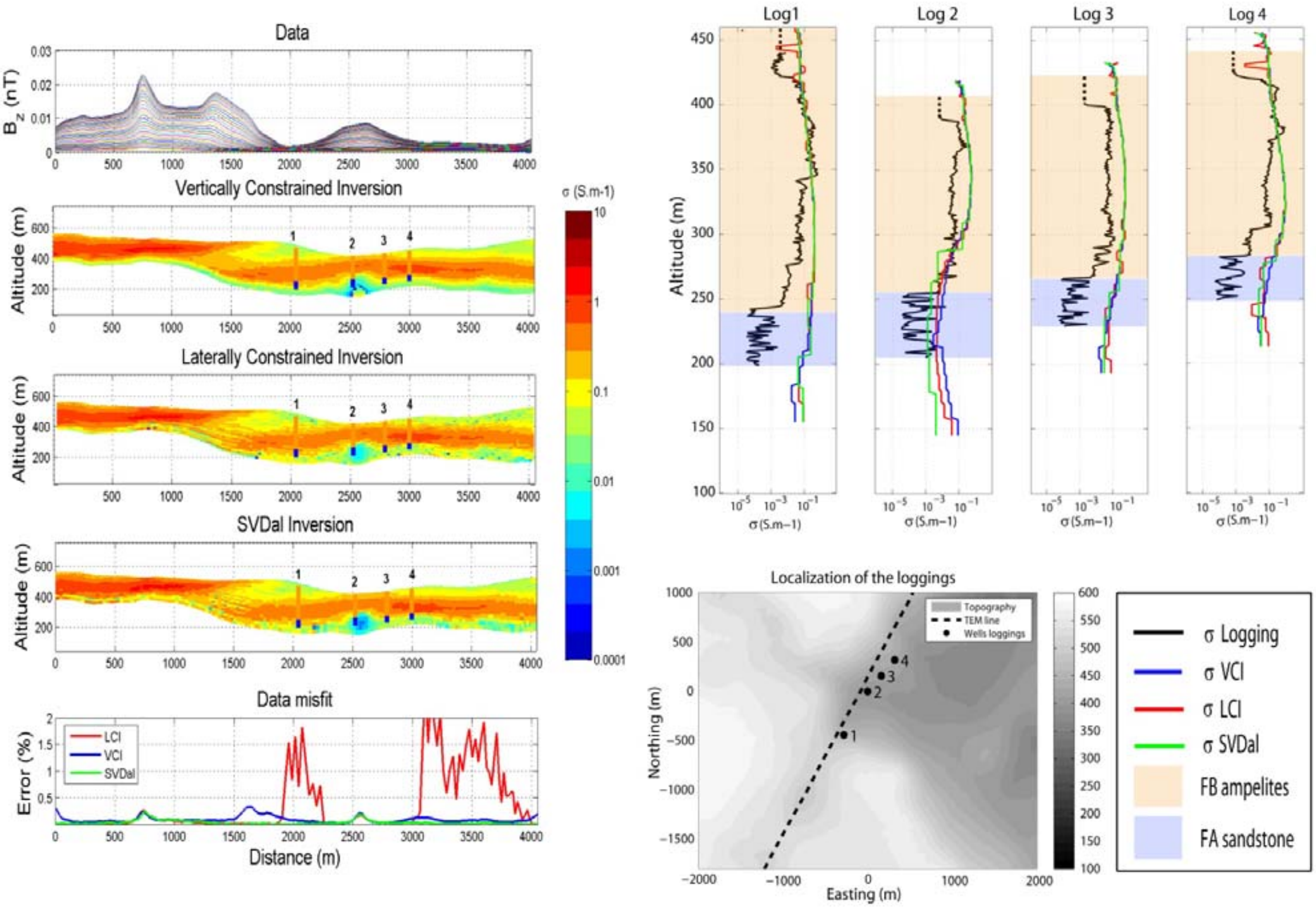

**Figure 7:** *Layered inversion of real data set acquired over the basin of Franceville (Gabon). On the left part: resulting conductivity sections for the three methods VCI, LCI and SVDal with additional logarithmic inversion. The lithologies FB and FA are superposed to the TEM section at wells positions in the profile. On the upper right part, the results of the three methods of TEM inversion are superposed to the drill holes measurement.*

By this way, the parameter $\lambda_p$ controls the magnitude of the perturbation used at each step. Thus, the electrical conductivity of the layered media is deduced using the following formula:

$$\boldsymbol{\sigma} = \boldsymbol{\sigma}^{\mathbf{0}} * e^{[\mathbf{F}^{*\mathbf{T}}\mathbf{F}^{*}+\lambda_{\mathrm{P}}{}^{2}\mathbf{I}]^{-\mathbf{1}}\mathbf{F}^{*\mathbf{T}}\mathbf{y}_{\mathbf{a}}} \; . \qquad (17)$$

If $\boldsymbol{\sigma}^{\mathbf{0}}$ is taken equal to the absolute value of the results provided by inversion of the conductivity, the logarithmic inversion can be used as an additional recursive filter which provides a realistic distribution of conductivity.

Figure 5 shows the application of this method on the synthetic data set. The starting model is the result of a vertical constrained inversion in which $\lambda_V$ has been chosen using the discrepancy principle for one sounding in the profile (this sounding is taken randomly). The magnitude of perturbation $\lambda_P$ has been selected in order to provide a good convergence of the data misfit.

**Limitation of the 1D interpretation**

As expected, these applications show that 1D layered inversion works quite well when imaging tabular conductive target with a low conductivity contrast. For this reason, 1D interpretation is relatively well adapted when applied to characterize most hydrological target. However, this approach fails to image properly local high conductivity contrasts which are typically encountered in mining exploration. Let us demonstrate this limitation by showing two synthetic cases with problematic results of the 1D inversion; synthetic data have been computed by 2.5 modelling using ArjunAir:

1- the first case is the end-border of a horizontally shaped formation with conductivity larger than the host,
2- the second case is vertically shaped like a dyke formation with conductivity larger than the host.

The conductivity of both targets is set equal to $\sigma_2 = 0.2$ S/m. The inversion is applied for two hosting media $\sigma_1 = 10^{-2}$ S/m and $\sigma_1 = 10^{-3}$ S/m. The results (Figure 6) show fake structures which seem to be dipping conductive slabs: this clearly illustrates how 1D interpretation is limited in cases of high conductivity contrast or vertically shaped formations. Consequently, more time-consuming method as 2D or 3D inversion with different regularization strategy (Portniaguine and Zhdanov, 1999) would be a more appropriate way to image these kinds of structures. This is also the conclusions of Ley-Cooper *et al.* (2010) who illustrated the limits of the LCI on synthetic 2D structures.

**Application to a real data set**

Let us consider the application of the layered inversion using these regularization methods on real helicopter-borne TEM data set (VTEM) acquired over the basin of Franceville in Gabon for mining exploration. We interpret the vertical component of the magnetic field which consists in 1500 soundings of 27 channels starting from $t_1 = 83\,\mu s$ to $t_2 = 7.8$ ms and acquired every 5m. The transmitter is a four turns loop with a diameter of 26 m. The electrical current of 200 A is injected during 8.32 ms before it is turned off; the pulse repetition rate is 25 Hz. The apparent conductivity is computed by table look up of pure step responses convolved with the measured waveform.

The basin is made of Precambrian sediments which can host mineralization of uranium. Usually, mineralization areas are found at the contact between two horizontal lithologies in the basin: FA sandstone and FB ampelites where uranium in solution has been precipitated thanks to the presence of organic matter. FB lithology is characterized by relatively high electrical conductivity that constitutes the top of a proterozoic reservoir and superposes on FA formation that is, composed of coarse grain size sediments characterized by relatively low electrical conductivity. This developed contrasts of conductivity of about three orders of magnitude which are often located at depths less than 400 m; therefore it is possible to detect it by using TEM imaging. The geology of the area has a tabular geometry with low dips that justifies the use of layered inversion as a first realistic approximation.

Robustness of the methods is clear on Figure 7 where we show the comparison between the conductivity obtained from the layered inversion of an airborne TEM line and the conductivity measured from logging into four boreholes in the same area. The topographic map at the bottom right corner of figure 7 shows the locations of boreholes and the TEM line considered. On Figure 7, the geological logs for each borehole are plotted on TEM conductivity sections. They are also displayed as background on the right part of the figure where TEM and borehole conductivity are compared.

Whatever the 1D TEM conductivities, it seems that the results are slightly more conductive than the conductivity measured in borehole (up to a factor of 2 within the first hundred meter depths). Similar considerations are found when comparing borehole conductivity with other kinds of EM data like CSEM. In case of TEM, this result may be due to the fact that TEM eddy currents are mostly horizontals while borehole conductivities are measured with vertical current lines. Tabular media are characterized by vertical transverse isotropy which shows larger conductivity in the horizontal directions than in the vertical direction. Therefore, TEM soundings, which are sensitive to the horizontal conductivity, should produce larger conductivity than borehole measurements. In addition, borehole conductivity is measured over a few centimetres samples of ground while TEM soundings integrate a larger area of several meters. This scaling change can generate non negligible differences between the two methods. Nevertheless, in a qualitative point of view, borehole 2 exhibits very good accordance between the different outcomes. The results at borehole 3 and 4 are coherent as well, except that we over-estimate the conductivity of the second layer in the TEM results (to factor of 10-20). The two latter boreholes are situated in the slope which separates the shelf from the valley. Since the shelf is supposed to have a thicker conductive layer on its surface, we suggest that the over-estimation of the second layer is the consequence of the topography which causes bias in the 1D TEM interpretations. We think that this effect occurs pathologically for borehole 1 which is situated at the top of the shelf border. Indeed, the TEM sounding does not detect the contact between the two lithologies.

**Conclusion**

All three effective ways which we have exposed to invert the electrical conductivity of a layered medium allow fast interpretation of a large volume of data during the survey. They make the regularization parameters more readable for geophysicists by using physical considerations as much as possible. Besides, it is important to note that all the different approaches provide similar images of conductivity. Therefore, one can think that the present methods should be used more as tools to avoid the mathematical instability due to data uncertainties than specific ways to manage the theoretical non-uniqueness (due to equivalent models).

The first method is the One Pass Imaging developed by Christensen (2002) in which we have simplified the parameterization by removing the term constraining the length of the solution. The second method is based on lateral constraints; it can be considered as a unidirectional constrained inversion since the *a priori* information comes from the previous sounding only and not from a set of surrounding ones. The major practical difference with those already reported in the literature is that soundings are inverted one by one; as a consequence, the computational cost is reduced. The third method allows the user to provide a model of conductivity which does not contain additional information given by regularization. The user still has to set one parameter to the program, the maximum value of the condition number.

Besides, we propose an iterative method which handles the problem with the logarithm of the conductivity; it is an additional step which starts from the result given by one of the three processes described previously. This method avoids the presence of negative values in the resulting conductivity model and lends realism to the conductivity distribution. Especially in case of tabular lithologies, resulting models show good accordance with the true model and the borehole measurements when applied on synthetic and real data respectively.

Not only does this method provide good accuracy as shown by evaluation on synthetic dataset and real VTEM survey, but it is really cost effective. We have written a dual C#/Matlab compiled program; it allows one to obtain a conductivity section resulting from 15000 TEM soundings (over a line of around 40 km for VTEM survey) in less than one minute on a Dual Core E8500 (3.16 GHz) with 3.25 Go of RAM.

We suggest that the method shown here for airborne TEM data inversion would be useful in the interpretation of other kinds of EM data, similarly in spectral EM and ground-based TEM but also in CSEM and MT where the layered medium is often used as a first approximation. Nevertheless, 2D or 3D inversion is necessary for imaging targets with sharp horizontal boundaries (e.g. faults and dykes); similar approach for the 2D or 3D inversion of apparent conductivity and regularization strategies can be used for fast airborne TEM imaging; this will be the topic of another paper.

**Acknowledgement**

This paper results from a Joint Research Project between AREVA-NC and CNRS UMR 7516. The authors thank AREVA Mining Business Group for financial support to J.G. as PhD student and for providing VTEM and borehole data.

## List of references